%% file: trans-cond-dp_v14.tex
\documentclass[11pt]{article}

\usepackage[top=1.0in,right=1.0in,left=1.0in,bottom=1.75in]{geometry}
\usepackage{amssymb,amsbsy}
\usepackage{amsmath}
\usepackage{amsthm}
\usepackage{graphics,graphicx}
\usepackage[T1]{fontenc}
\usepackage{color}
\usepackage{subfigure}
\usepackage[affil-it,auth-sc]{authblk}
\usepackage{stmaryrd}
\usepackage{booktabs}
\usepackage{array}

\usepackage[]{jabbrv}

\usepackage[numbers,sort&compress]{natbib}

\newcolumntype{L}[1]{>{\raggedright\let\newline\\\arraybackslash\hspace{0pt}}m{#1}}
\newcolumntype{C}[1]{>{\centering\let\newline\\\arraybackslash\hspace{0pt}}m{#1}}
\newcolumntype{R}[1]{>{\raggedleft\let\newline\\\arraybackslash\hspace{0pt}}m{#1}}

\input{definitions}

\newcommand{\ljump}{\llbracket}
\newcommand{\rjump}{\rrbracket}
\newcommand{\jump}[1]{\ljump #1 \rjump}

\graphicspath{{./}{./figures/}}

\title{General transmission conditions for \\ thin elasto-plastic pressure-dependent interphase \\
between dissimilar materials}

\date{}

\author[1]{M. Sonato}
\author[1]{A. Piccolroaz\footnote{Corresponding author: e-mail: roaz@ing.unitn.it; phone: +39\,0461\,282583.}}
\author[2]{W. Miszuris}
\author[2]{G. Mishuris}

\affil[1]{Department of Civil, Environmental and Mechanical Engineering, University of Trento, Italy}
\affil[2]{Department of Mathematics, Aberystwyth University, Aberystwyth, U.K.}

\begin{document}

\maketitle

\begin{abstract}
\noindent
We consider a thin soft adhesive interphase between dissimilar elastic media. The material of the intermediate layer is modelled by elasto-plastic pressure-sensitive constitutive law. An asymptotic procedure, together with a novel formulation of the deformation theory of plasticity for pressure-sensitive materials, is used in order to derive nonlinear transmission conditions for the corresponding imperfect zero-thickness interface. A FEM analysis of the original three-phase structure is performed to validate the transmission conditions for the simplified bimaterial structure.
\end{abstract}


%
%

\section{Introduction}
\label{sec01}

Adhesive joints are widely used in structural engineering applications involving composite materials, such as marine \cite{Golaz2013}, aerospace \cite{Bhowmik2009} and automotive \cite{Loureiro2010765} structures. In these structures, polymeric adhesives are used to join different materials such as plastics, metals, fibre reinforced composites and others.

Early models of adhesive joints tended to assume that the adhesive behaves as a linear-elastic solid \cite{Bikerman1961,Adams1974}. However, many adhesives (e.g. rubber modified epoxies) exhibit large plastic strains before failure \cite{Wang2000}. A number of yield criteria have been proposed to model the plastic behaviour of polymers. Among these, the Tresca and von Mises criteria, originally developed to describe the yield behaviour of metals, have been also used. However, these criteria cannot accurately predict the behaviour of polymeric adhesive specimens under multiaxial loading, as yielding in these materials is sensitive to hydrostatic as well as the deviatoric stress \cite{Rabinowitz1970,Altenbach2001}. As a consequence, pressure-dependent plasticity theory is more appropriate in this case. In this context, the Drucker-Prager yield criterion is widely used for polymeric materials.

The numerical FE modelling of thin adhesive interphases is complicated by the very large aspect ratio of the intermediate layer, which requires extremely refined meshes. Alternatively, the so-called imperfect interface approach can be adopted. It consists of replacing the actual thin interphase between the two adherents by a zero thickness imperfect interface. The interface is supplied with special transmission conditions, which incorporate information about geometrical and mechanical properties of the original thin interphase.

Although imperfect transmission conditions for elastic interphases have been intensively investigated \cite{Benveniste1985,Benveniste2001,Avila-Pozos1999,Hashin2002}, analytical results for elastoplastic adhesive interphase are very scarce and mainly limited to rigid adherents and pressure-independent plasticity \cite{Ikeda2000,Mishuris2007}.

In this paper, nonlinear transmission conditions for a soft elasto-plastic {\it pressure-dependent} interphase are derived by asymptotic techniques. The interphase material is described by a novel formulation of the deformation theory for pressure-dependent media. Early models of pressure-dependent deformation theory can be found in \cite{Durban2003} and \cite{chapter}. However, the model proposed by \cite{Durban2003} only accounts for elastic and plastic branches separately, i.e. the yield condition is not incorporated into the generalized constitutive parameters. On the other hand, the model proposed in \cite{chapter} applies only to a few specific loading paths. To overcome these limitations, the classical Hencky deformation theory is generalized in Section \ref{sec03} to the case of pressure-sensitive materials by including the dependence of inelastic deformation on the first stress invariant, in the spirit suggested by Chen \cite{Chen1988} and Lubarda \cite{Lubarda2000}. The general theory is then specialized to the  the case of Drucker-Prager model with associated flow rule and isotropic linear hardening.

The transmission conditions are validated by numerical examples based on accurate finite element simulations. The accuracy and efficiency of the proposed
approach prove to be very high for different monotonic loading conditions.

\section{Problem formulation and transmission conditions for elasto-plastic pressure-dependent interphases}
\label{sec02}

\begin{figure}[!htcb]
\centering
\includegraphics[width=90mm]{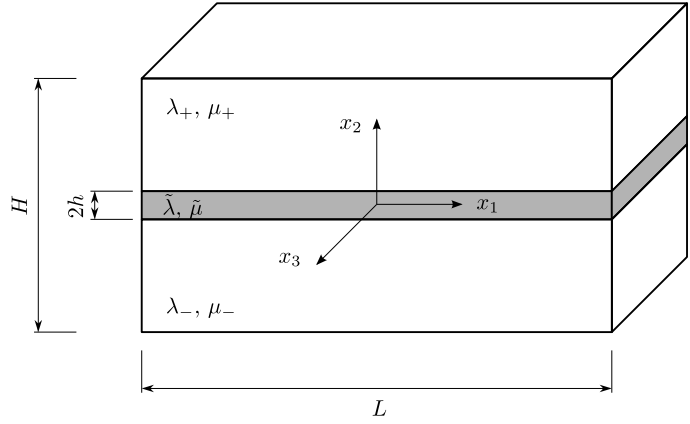}
\caption{Bimaterial structure with thin soft adhesive joint.}
\label{fig00}
\end{figure}

Consider a structure composed by two dissimilar elastic materials joined together by a thin adhesive interphase, see Fig.~\ref{fig00}. The thickness of the
interphase is small in comparison with the characteristic size of the body: $h = \epsilon h_* \ll L$, $h_* \sim L$. Here $\epsilon$ is a small positive parameter,
$\epsilon \ll 1$.

The adhesive material is assumed to be soft in comparison with the two adherents and may exhibit a very general nonlinear constitutive behaviour, including
compressible plastic deformations, the only assumption being that the material is isotropic.

Adopting the deformation theory of plasticity, the constitutive laws of the interphase in the plastic regime will be described in terms of nonlinear elasticity
as
\begin{equation}
\label{hooke}
\sigma_{ij} = \tilde{\lambda} \varepsilon_{kk} \delta_{ij} + 2 \tilde{\mu} \varepsilon_{ij},
\end{equation}
where the generalized Lam\'e parameters are functions depending on the deformation within the interphase
\begin{equation}
\label{lame}
\tilde{\lambda} = \tilde{\lambda}(J_1^\varepsilon,J_2^e), \quad \tilde{\mu} = \tilde{\mu}(J_1^\varepsilon,J_2^e).
\end{equation}
Note that, due to the isotropy assumption, the dependence is written in terms of the two invariants
\begin{equation}
J_1^\varepsilon = \varepsilon_{kk}, \quad J_2^e = \frac{1}{2} e_{ij} e_{ij},
\end{equation}
where $e_{ij}$ denotes the deviatoric part of the strain tensor. For the sake of keeping the analysis simple, the dependence on the third invariant is omitted.
However, if necessary, the model can be generalized to include the effect of all invariants.

In the elastic regime, the elastic constants are assumed to be constant
\begin{equation}
\lambda = \epsilon \lambda_0, \quad \mu = \epsilon \mu_0,
\end{equation}
where the elastic parameters $\lambda_0$, $\mu_0$ are comparable with those of the adherent materials, $\lambda_\pm$, $\mu_\pm$.

We assume in our analysis that, for any admissible deformation of the material within the interphase, there exists a constant parameter $\nu_*$ such that
\begin{equation}
0 \leq \tilde{\nu}(J_1^\varepsilon,J_2^e) \leq \nu_* < 1/2,
\end{equation}
where $\tilde{\nu}(J_1^\varepsilon,J_2^e)$ is the generalized Poisson's ratio
\begin{equation}
\tilde{\nu}(J_1^\varepsilon,J_2^e) = \frac{\tilde{\lambda}(J_1^\varepsilon,J_2^e)}{2[\tilde{\lambda}(J_1^\varepsilon,J_2^e) + \tilde{\mu}(J_1^\varepsilon,J_2^e)]}.
\end{equation}
Under such assumption, the following two imperfect transmission conditions linking the values from different parts of the interphase are asymptotically satisfied,
see \cite{chapter}
\begin{equation}
\label{tc1}
\jump{\bt_2}(x_1,x_3) = 0,
\end{equation}
and
\begin{equation}
\label{tc2}
\bt_2(x_1,x_3) = \frac{1}{2h} \bA(x_1,x_3) \jump{\bu}(x_1,x_3),
\end{equation}
where $\bt_2 = [\sigma_{21},\sigma_{22},\sigma_{23}]^T$ is the traction on the interphase surfaces, $\bu = [u_1,u_2,u_3]^T$ is the displacement vector, and the
jump of a function $f$ across the interphase is denoted by $\jump{f} = f(x_2 = h) - f(x_2 = -h)$, and the tensor $\bA$ is given by
\begin{equation}
\bA =
\begin{pmatrix}
\tilde{\mu} & 0 & 0 \\
0 & \tilde{\lambda} + 2\tilde{\mu} & 0 \\
0 & 0 & \tilde{\mu}
\end{pmatrix}.
\end{equation}
The transmission conditions (\ref{tc1}) and (\ref{tc2}) are valid under the following ellipticity conditions
\begin{equation}
\label{ass}
0 < \tilde{\mu} < \epsilon \mu_*, \quad 0 < \tilde{\lambda} + 2\tilde{\mu} < \epsilon \Lambda_*,
\end{equation}
where $\mu_*$ and $\Lambda_*$ are constants comparable in values with those of the adherent materials. In fact, we can take $\mu_* = \mu_0$ and
$\Lambda_* = \lambda_0 + 2\mu_0$.

The asymptotic procedure allows us also to estimate the components of the strain tensor, which to the leading order $O(\epsilon^{-1})$ read
\begin{equation}
\varepsilon_{ij} = O(1), \quad \varepsilon_{i2} = \frac{1}{2} \frac{\partial u_i}{\partial x_2} + O(1), \quad i,j = 1,3, \quad
\varepsilon_{22} = \frac{\partial u_2}{\partial x_2}.
\end{equation}
By application of the generalized Hooke's law (\ref{hooke}), we obtain the following relationships
\begin{equation}
\varepsilon_{i2} = \frac{\sigma_{i2}}{2\tilde{\mu}}, \quad i,j = 1,3, \quad \varepsilon_{22} = \frac{\sigma_{22}}{\tilde{\lambda} + 2\tilde{\mu}},
\end{equation}
Finally, the leading terms of the strain tensor invariants can be calculated as
\begin{equation}
\label{j1}
J_1^\varepsilon = \varepsilon_{22} = \frac{\sigma_{22}}{\tilde{\lambda} + 2\tilde{\mu}},
\end{equation}
\begin{equation}
\label{j2}
J_2^e = \frac{1}{3} \varepsilon_{22}^2 + \varepsilon_{12}^2 + \varepsilon_{23}^2 =
\frac{1}{3} \frac{\sigma_{22}^2}{(\tilde{\lambda} + 2\tilde{\mu})^2} + \frac{1}{4} \frac{\sigma_{12}^2}{\tilde{\mu}^2} + \frac{1}{4} \frac{\sigma_{23}^2}{\tilde{\mu}^2}.
\end{equation}
Note that the material functions $\tilde{\lambda}$ and $\tilde{\mu}$, at this point, are not specified and should be defined depending on the particular
elasto-plastic model adopted for the interphase material. However, all the stress components $\sigma_{i2}$ of the traction vector $\bt_2$ as well as the
components of the strain tensor $\varepsilon_{ij}$ in their leading terms are functions only of the space variables $x_1$ and $x_3$. This immediately implies
that, in the case under consideration, the strain invariants are also functions of those two variables,
\begin{equation}
J_1^\varepsilon = J_1^\varepsilon(x_1,x_3), \quad J_2^e = J_2^e(x_1,x_3).
\end{equation}
The transmission conditions (\ref{tc2}) are then used to write the strain invariants (\ref{j1}) and (\ref{j2}) in terms of components of displacement jump
\begin{equation}
\label{fin1}
J_1^\varepsilon = \frac{1}{2h} \jump{u_2},
\end{equation}
\begin{equation}
\label{fin2}
J_2^e = \frac{1}{12h^2} \jump{u_2}^2 + \frac{1}{16h^2} \jump{u_1}^2 + \frac{1}{16h^2} \jump{u_3}^2.
\end{equation}
This allows us to formulate the nonlinear transmission conditions for a thin soft nonlinear interphase in their final form
\begin{equation}
\label{trcond}
\jump{\bt_2} = 0, \quad \bt_2 = \frac{1}{2h} \bA(\jump{\bu})\jump{\bu},
\end{equation}
where the matrix-function $\bA$ is explicitly defined by $\bA(\jump{\bu}) = \bA(\tilde{\lambda}(J_1^\varepsilon,J_2^e),\tilde{\mu}(J_1^\varepsilon,J_2^e))$ and the relationships (\ref{fin1}) and (\ref{fin2}).

Let us recall that the assumptions (\ref{ass}) should be satisfied to ensure the applicability of the transmission conditions. However, even if the estimates (\ref{ass}) cannot be checked a priori, the nonlinear transmission conditions (\ref{trcond}) can be successfully used in the modelling. In this case, a posteriori verification of (\ref{ass}) will be necessary in order to confirm the validity of the final results.

Summarizing, eqs. (\ref{trcond}) constitute the transmission conditions describing a soft imperfect interface in a bimaterial structure. These conditions allows one to replace the thin plastic interphase with an imperfect nonlinear interface (of zero thickness) and consider a simplified outer problem for the two bonded materials. The thin interphase influences the solution only via the imperfect transmission conditions which incorporates the most important information on the properties of the adhesive material and the nonlinear deformations developing within the interphase.

However, in order to use the nonlinear imperfect transmission conditions (\ref{trcond}) one needs to define the generalized Lam\'e parameters (\ref{lame}) as functions of the strain invariants depending on the mechanical properties of the interphase material. If the parameters are constant, the conditions reduces to the thin soft elastic interphase case. In case of elasto-plastic material obeying the von Mises criterion, the analysis was presented in \cite{Mishuris2007}. In the next section, we derive the proper deformation theory for a pressure-dependent plastic material, with particular reference to the Drucker-Prager criterion.

\section{Pressure-dependent deformation theory}
\label{sec03}

\subsection{Main Assumptions}
\label{sec0301}

In this section, the classical $J_2$ deformation theory is generalized to the case of pressure-sensitive materials, so that the first invariant $J_1$ will also be included in the formulation. We make the following assumptions:
\begin{itemize}
\item The material is initially isotropic.
\item The principal axes of the plastic strain tensor $\varepsilon_{ij}^p$ are coincident with those of the stress tensor $\sigma_{ij}$.
\item The plastic deviatoric strain tensor $e_{ij}^p$ is proportional to the deviatoric stress tensor $s_{ij}$.
\end{itemize}
In the following, it is convenient to introduce the volumetric and deviatoric decomposition of the stress and strain tensor as:
\begin{align}
\sigma_{ij} &=  \sigma_m\delta_{ij} + s_{ij}\,, \quad \sigma_m = \frac{1}{3} \sigma_{kk}, \label{decomp_sigma} \\
\varepsilon_{ij} &= \varepsilon_m\delta_{ij} + e_{ij}\,, \quad \varepsilon_m = \frac{1}{3} \varepsilon_{kk}, \label{decomp_eps}
\end{align}
and the additive composition of the elastic and plastic strains
\begin{equation}
\label{add_comp}
\varepsilon_{ij} = \varepsilon_{ij}^e + \varepsilon_{ij}^p =
\left(\varepsilon_m^e + \varepsilon_m^p\right)\delta_{ij} + e_{ij}^e + e_{ij}^p,
\end{equation}
in order to derive the closed-form relationship. According to assumptions~(2) and (3), the plastic strain can be written as
\begin{equation}
\label{e_phi_s}
\varepsilon_{ij}^p = \phi_{1} \sigma_m \delta_{ij} + \phi_{2} s_{ij},
\end{equation}
where $\phi_{1}$ and $\phi_{2}$ are functions which represent the hardening behaviour of the material. In the case of plastic loading, it holds that $\phi_{1}\ne 0$ and $\phi_{2}> 0$, while in the case of elastic loading $\phi_{1} = \phi_{2} = 0 $ prevails. Note that Eq.~(\ref{e_phi_s}) can be derived following Chen's suggestions \cite{Chen1988}.

The constitutive relation between stress and elastic strain decomposed in its volumetric and deviatoric parts is given by
\begin{equation}
e_{ij}^e = \cfrac{s_{ij}}{2G}, \quad
\varepsilon_m^e = \cfrac{\sigma_m}{3K}.
\end{equation}
The material is assumed to be elastic compressible, so that $0 \leq \nu < 1/2$ and $K < \infty$.

Under the given assumptions, the nonlinear stress-strain law for a pressure-dependent plastic material becomes
\begin{equation}
\label{Hencky_stress_strain}
\sigma_{ij} = 3\tilde{\lambda}\,\varepsilon_m \delta_{ij} + 2\tilde{\mu}\, \varepsilon_{ij},
\end{equation}
where the generalized Lam\'e's coefficients have been introduced:
\begin{equation}
\label{t_lambda}
\tilde{\lambda}(\phi_{1},\phi_{2}) = \cfrac{3\nu + (\phi_{2} - \phi_{1})E}{3(1 + \nu + \phi_{2} E)(1 - 2\nu + \phi_{1} E)} \, E, \quad
\tilde{\mu}(\phi_{2}) = \cfrac{E}{2(1 + \nu + \phi_{2} E)}, 
\end{equation}
It should be noted here that these coefficients coincide in the pure elastic regime $(\phi_{1} = \phi_{2} = 0)$ with the elastic Lam\'{e}'s coefficients. Further
relationships for generalized elastic constants can be derived in the same manner:
\begin{equation}
\label{const}
\tilde{E}(\phi_{1},\phi_{2}) = \cfrac{3E}{3 + (2\phi_{2} + \phi_{1})E}, \quad
\tilde{\nu}(\phi_{1},\phi_{2}) = \cfrac{3\nu + (\phi_{2} - \phi_{1})E}{3 + (2\phi_{2} + \phi_{1})E}, \quad
\tilde{K}(\phi_{1}) = \cfrac{E}{3(1 - 2\nu + \phi_{1} E)}.
\end{equation}

To define the functions $\phi_{1}$ and $\phi_{2}$, let us consider the relationships for the first and second invariants of the strain and stress tensor, i.e.
\begin{equation}
J_1^{\varepsilon^p} = \varepsilon^p_{kk},  \quad J_1^{\sigma} = \sigma_{kk}, \quad 
J_2^{e^p} = \cfrac{1}{2} e^p_{ij}e^p_{ij}, \quad J_2^{s} = \cfrac{1}{2} s_{ij}s_{ij},
\end{equation}
where the proportionality relation \eqref{e_phi_s} can be used to obtain:
\begin{equation}
J_1^{\varepsilon^p} = 3\phi_{1}\sigma_m = \phi_{1} J_1^\sigma, \quad
J_2^{e^p} = \cfrac{1}{2}\phi_{2}^2 s_{ij}s_{ij} = \phi_{2}^2 J_2^s.
\end{equation}
Thus, the scalar functions $\phi_{1}$ and $\phi_{2}$ can be obtained in the case of a multiaxial stress state as
\begin{equation}
\label{f1}
\phi_{1} = \cfrac{J_1^{\varepsilon^p}}{J_1^\sigma}, \quad
\phi_{2} = \cfrac{\sqrt{J_2^{e^p}}}{\sqrt{J_2^s}}.
\end{equation}
In order that these relations can be used in the analysis of the thin plastic layer, the functions $\phi_1$ and $\phi_2$ must be defined in terms of total strain invariants. From the nonlinear elasticity relationship (\ref{Hencky_stress_strain}), it is noted that the first invariants of the stress tensor and the total strain tensor are linearly dependent,
\begin{equation}
\label{first_inv}
J_1^\sigma = (3\tilde\lambda + 2\tilde \mu)\, J_1^{\varepsilon} = 3 \tilde{K}\, J_1^{\varepsilon},
\end{equation}
whereas the second invariants of the deviatoric tensors are related by
\begin{equation}
\label{second_inv}
\sqrt{J_2^s} = 2 \tilde{\mu}\, \sqrt{J_2^e}.
\end{equation}
This, in turns, allows us to evaluate the generalized material parameters as functions of the invariants
\begin{equation}
\label{functions_inv}
2\tilde{\mu} = \frac{\sqrt{J_2^s}}{\sqrt{J_2^e}}, \quad
3\tilde{K} = \frac{J_1^\sigma}{J_1^\varepsilon}.
\end{equation}
Substituting these relations into Eqs.~(\ref{t_lambda}) and (\ref{const}), one can define after some algebra:
\begin{equation}
\label{phi}
\phi_{1} = \cfrac{J_1^{\varepsilon}}{J_1^\sigma} - \frac{1 - 2\nu}{E}, \quad
\phi_{2} = \cfrac{\sqrt{J_2^{e}}}{\sqrt{J_2^s}} - \frac{1 + \nu}{E}.
\end{equation}
These relationships are more useful in the analysis than (\ref{f1}). This also gives the well known relationships between the invariants of the plastic strain and the total strain\footnote{
The last formula can be also justified by the fact that the tensors $\be^e$, $\be^p$ and $\be$ are proportional to each other:
$$
\be^p = 2\mu \phi_2 \be^e = \cfrac{1}{1 + 2\mu \phi_2}\, \be.
$$
}:
\begin{equation}
\label{general}
J_1^{\varepsilon} = J_1^{\varepsilon^p} + \frac{1 - 2\nu}{E} J_1^\sigma, \quad
\sqrt{J_2^{e}} = \sqrt{J_2^{e^p}} + \frac{1 + \nu}{E} \sqrt{J_2^s}.
\end{equation}
In the next section, we develop the deformation theory in the case of uniaxial stress state. In Section \ref{sec0303}, the deformation theory is generalized to the case of general multiaxial state and Drucker-Prager material.

\subsection{Uniaxial Stress Test}
\label{sec0302}

\begin{figure}[!htcb]
\centering
\includegraphics[width=150mm]{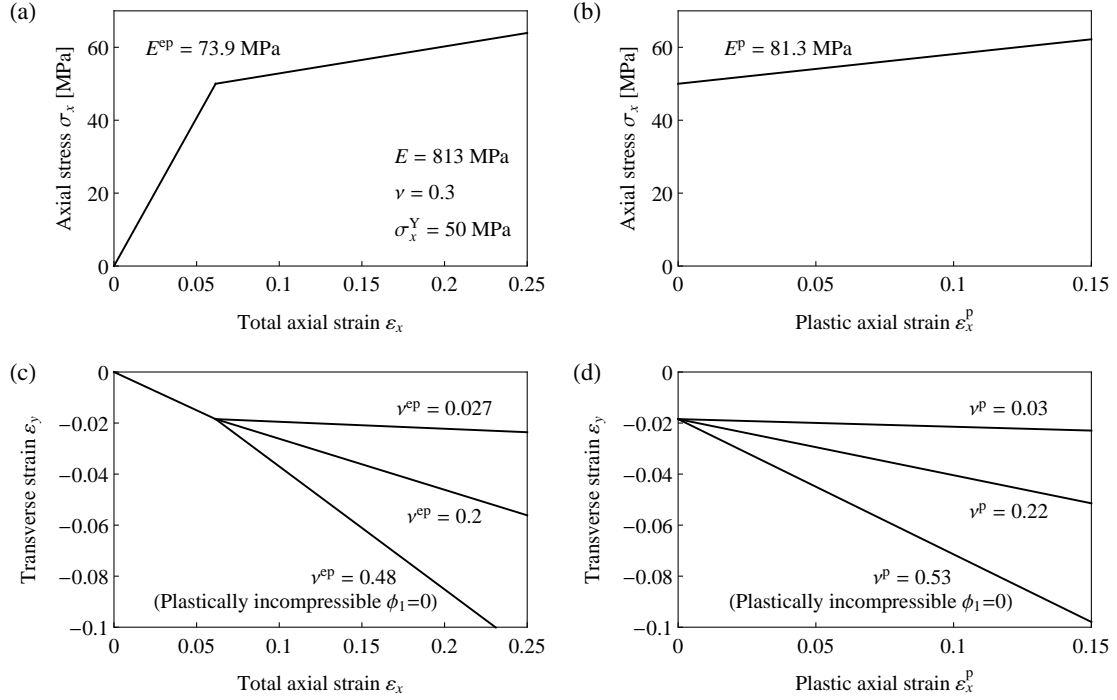}
\caption{Example of constitutive behaviour in a uniaxial stress test for a linear hardening material: (a) Axial stress $\sigma_x$ as a function of the total axial strain $\varepsilon_x$; (b) Axial stress $\sigma_x$ as a function of the plastic axial strain $\varepsilon_x^p$; (c) Transverse contraction $\varepsilon_y$ as a function of the total axial strain $\varepsilon_x$; (d) Transverse contraction $\varepsilon_y$ as a function of the plastic axial strain $\varepsilon_x^p$. Three cases of transverse contraction are considered: plastically incompressible material $\varepsilon_m^p = 0$ ($\nu^{ep} = 0.48$ and $\nu^p = 0.53$), null plastic transverse contraction $\varepsilon_y^p = 0$ ($\nu^{ep} = 0.027$ and $\nu^p = 0.03$) and an intermediate case ($\nu^{ep} = 0.2$ and $\nu^p = 0.22$).}
\label{fig01}
\end{figure}

The uniaxial stress test is one of the most common tests, widely accessible, and it may provide at least two sets of experimental data: $\sigma_x(\varepsilon_x)$ and $\varepsilon_y(\varepsilon_x)$.
The specific form of the scalar functions $\phi_{1}$ and $\phi_{2}$ for the case of a uniaxial stress state can be derived in the following way. Assuming that the direction of loading coincides with the $x$-direction $(\sigma_x \neq 0; \varepsilon_y^p = \varepsilon_z^p)$, the scalar functions $\phi_1$ and $\phi_2$ become
\begin{equation}
\phi_{1} = \cfrac{\varepsilon_x^p + 2\varepsilon_y^p}{\sigma_{x}}, \quad
\phi_{2} = \cfrac{|\varepsilon_x^p - \varepsilon_y^p|}{\sigma_{x}}.
\end{equation}
Let us assume in the following an elastic-plastic material with linear hardening as shown in Fig.~\ref{fig01}.
%

\begin{figure}[!htcb]
\centering
\includegraphics[width=150mm]{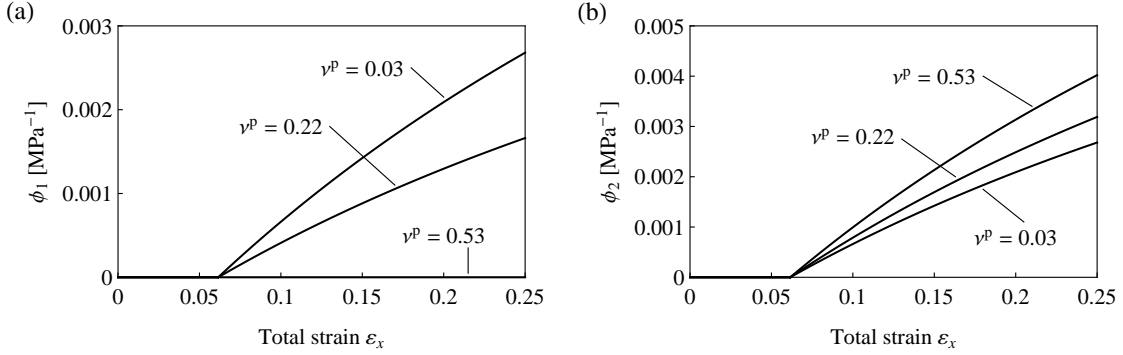}
\caption{(a) Function $\phi_1(\varepsilon_x)$ and (b) function $\phi_2(\varepsilon_x)$ corresponding to the material parameters presented in Fig.~\ref{fig01}.}
\label{fig02}
\end{figure}

\noindent
The flow stress $\sigma_x$ can be expressed in terms of the plastic modulus $E^p$ as (cf.\ Fig.~\ref{fig01}b)
\begin{equation}
\label{sigx}
\sigma_x(\varepsilon_x^p) = E^p \varepsilon_x^p + \sigma_s,
\end{equation}
whereas the transversal deformation can be expressed as (cf.\ Fig.~\ref{fig01}d)
\begin{equation}
\varepsilon_y(\varepsilon_x^p) =
-\nu^p \varepsilon_x^p - \cfrac{\nu}{E} \sigma_s.
\end{equation}
The plastic part of $\varepsilon_y$ can be derived as
\begin{equation}
\label{contract}
\varepsilon_y^p =
\varepsilon_y - \varepsilon_y^e =
-\nu^p \varepsilon_x^p - \cfrac{\nu}{E} \sigma_s + \nu \cfrac{\sigma_x}{E} =
-\left( \nu^p - \nu \cfrac{E^p}{E} \right) \varepsilon_x^p,
\end{equation}
so that the functions $\phi_{1}$ and $\phi_{2}$ are obtained in terms of the plastic strain in a uniaxial stress state as
\begin{equation}
\phi_{1}(\varepsilon_x^p) =
\cfrac{(E - 2\nu^pE + 2\nu E^p)\varepsilon_x^p}{E(E^p \varepsilon_x^p + \sigma_s)}, \quad
\phi_{2}(\varepsilon_x^p) =
\cfrac{(E + \nu^pE - \nu E^p)\varepsilon_x^p}{E(E^p \varepsilon_x^p + \sigma_s)}.
\end{equation}
Using the additive composition of the elastic and plastic strains, i.\,e. $\varepsilon_x = \varepsilon_x^e + \varepsilon_x^p$, the stress $\sigma_x$ and the transversal deformation $\varepsilon_x^p$ are obtained as functions of the total strain $\varepsilon_x$
\begin{equation}
\sigma_x(\varepsilon_x) = E^{ep} \varepsilon_x + \cfrac{E - E^{ep}}{E} \sigma_s, \quad
\varepsilon_y(\varepsilon_x) = - \nu^{ep} \varepsilon_x - \cfrac{\nu - \nu^{ep}}{E} \sigma_s,
\end{equation}
where
\begin{equation}
E^{ep} = \cfrac{E E^p}{E + E^p}, \quad \nu^{ep} = \cfrac{\nu^p E}{E + E^p}.
\end{equation}
Finally, the dependence of $\phi_{1}$ and $\phi_{2}$ upon the total strain $\varepsilon_x$ is given by
\begin{align}
\label{eq:phi1}
\phi_{1}(\varepsilon_x) &= \cfrac{1}{E} \left(1 - 2\nu^p + 2\nu \cfrac{E^p}{E}\right)
\cfrac{E\varepsilon_x - \sigma_s}{E^p\varepsilon_x + \sigma_s}, \\
\phi_{2}(\varepsilon_x) &= \cfrac{1}{E} \left(1 + \nu^p - \nu \cfrac{E^p}{E}\right)
\cfrac{E\varepsilon_x - \sigma_s}{E^p\varepsilon_x + \sigma_s}. \label{eq:phi2}
\end{align}
The functions $\phi_1(\varepsilon_x)$ and $\phi_2(\varepsilon_x)$ corresponding to the material parameters presented in Fig.~\ref{fig01} are shown in Fig.~\ref{fig02}.
%
%
Interestingly, in the case of uniaxial stress state the ratio of the functions $\phi_{1}(\varepsilon_x)$ and $\phi_{2}(\varepsilon_x)$ is a constant value:
\begin{equation}
\label{ratio_fi}
\frac{\phi_{1}(\varepsilon_x)}{\phi_{2}(\varepsilon_x) }=\frac{(1 - 2\nu^p)E + 2\nu E^p}{(1 + \nu^p)E - \nu E^p}.
\end{equation}

\subsection{Pressure-dependent yielding: Drucker-Prager Material}
\label{sec0303}

Consider a material obeying the Drucker-Prager yield criterion,
\begin{equation}
\label{DP}
F(J_1^\sigma,J_2^s) = \alpha J_1^\sigma + \sqrt{J_2^s} - k_s,
\end{equation}
where $\alpha > 0$ is called pressure-sensitivity index and assumed to be a constant while $k_s$ is a positive function of the plastic strain
\begin{equation}
\label{k_s}
k_s = k_s(\bvarepsilon^p) > 0.
\end{equation}
In the case of uniaxial stress test
\begin{equation}
\label{hard}
k_s(\varepsilon_x^p) = \alpha J_1^\sigma + \sqrt{J_2^s} =
\left( \alpha + \cfrac{1}{\sqrt{3}} \right) \left( E^p \varepsilon_x^p + \sigma_s \right).
\end{equation}
Assuming the normality rule (plastic strain increment proportional to the normal of the yield surface $n_{ij} = 2\alpha\sqrt{J_2^s}\delta_{ij} + s_{ij}$)
and proportional loading we have (cf.~\cite{Lubarda2000})
\begin{equation}
\varepsilon^p_{ij} = \phi_2 (2\alpha\sqrt{J_2^s}\delta_{ij} + s_{ij}),
\end{equation}
so that
\begin{equation}
\label{j1j2}
J_1^{\varepsilon^p} = 6\alpha \phi_2 \sqrt{J_2^s}, \quad
\sqrt{J_2^{e^p}} = \phi_2 \sqrt{J_2^s}.
\end{equation}
Note that the ratio between the two plastic strain invariants is constant
\begin{equation}
\label{ratio}
\frac{J_1^{\varepsilon^p}}{\sqrt{J_2^{e^p}}} = 6\alpha.
\end{equation}
The general setting presented in Section \ref{sec0301} can be specialized to this type of material by defining
\begin{equation}
\label{D-P_1}
\phi_1 = \cfrac{6\alpha \sqrt{J_2^s} \phi_2}{J_1^\sigma}, \quad
\phi_2 = \phi_2.
\end{equation}
It is important to emphasize here that both functions $\phi_1$ and $\phi_2$ equal to zero simultaneously in the elastic regime or are both different from zero in the plastic zone.

We need to represent now the functions $\phi_1$ and $\phi_2$ depending only on the invariants of the total strain tensor. Let us assume that we have been
successful with the second function,
\begin{equation}
\label{general_fi}
\phi_2 = \phi_2(J_1^\varepsilon,J_2^e),
\end{equation}
then the remaining function $\phi_1$ can be computed from Eq.~(\ref{D-P_1}) under taking into account  Eqs.~(\ref{functions_inv}), (\ref{t_lambda}) and (\ref{const})
\begin{equation}
\phi_1 = 6\alpha \cfrac{\sqrt{J_2^s} \phi_2}{J_1^\sigma} =
6\alpha \phi_2 \frac{2\tilde\mu}{3\tilde K}\frac{\sqrt{J_2^e}}{J_1^\varepsilon} =
6\alpha \phi_2 \frac{1 - 2\nu + \phi_1 E}{1 + \nu + \phi_2 E}\frac{\sqrt{J_2^e}}{J_1^\varepsilon},
\end{equation}
or
\begin{equation}
\label{fi_1}
\phi_1 = \phi_1(J_1^\varepsilon,J_2^e) =
\cfrac{6\alpha (1 - 2\nu) \phi_2 \sqrt{J_2^e}}{J_1^\varepsilon(1 + \nu) + \phi_2 E\big(J_1^\varepsilon - 6\alpha \sqrt{J_2^e}\big)}.
\end{equation}

\begin{figure}[!htcb]
\centering
\includegraphics[width=160mm]{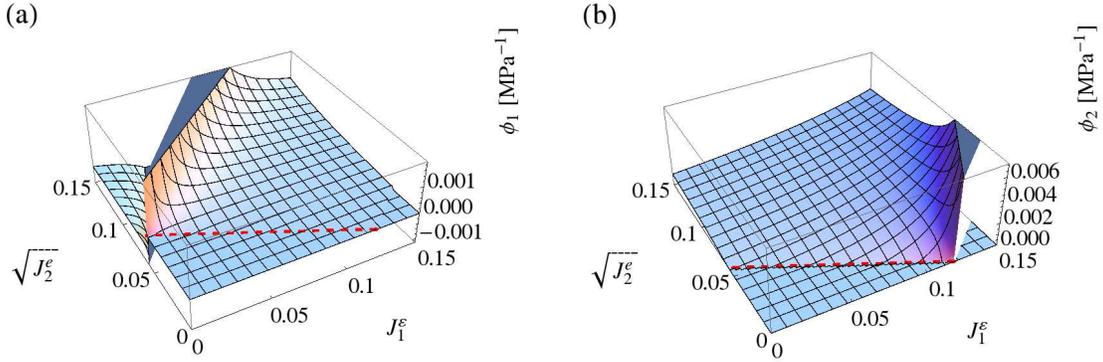}
\caption{(a) Function $\phi_1$ and (b) function $\phi_2$ of Drucker-Prager's material with linear hardening as functions of the two invariants of strain, $J_1^\varepsilon$
and $J_2^e$. The dashed red line indicates the elastic limit.}
\label{fig04}
\end{figure}

Using Eqs.~(\ref{general_fi}) and (\ref{fi_1}), the generalized elastic constants defined in Eqs.~(\ref{t_lambda}) and (\ref{const}) can be rewritten in terms of
the only two invariants of the total strain tensor:
\begin{align}
\label{const_D-P1}
\tilde{\mu}(\phi_2) &=
\tilde G(\phi_2) = \cfrac{E}{2(1 + \nu + \phi_2 E)}, \\
\tilde{E}(\phi_2) &=
\cfrac{3E}{1 + \nu + \phi_2 E} \frac{J_1^\varepsilon(1 + \nu + \phi_2 E) - 6\alpha\phi_2 E\sqrt{J_2^e}}{J_1^\varepsilon(3 + 2\phi_2 E) -
12\alpha\phi_2 E\sqrt{J_2^e}}, \\
\tilde{\nu}(\phi_2) &=
\frac{J_1^\varepsilon(3\nu + \phi_2 E) - 6\alpha\phi_2 E\sqrt{J_2^e}}{J_1^\varepsilon(3 + 2\phi_2 E) - 12\alpha\phi_2 E\sqrt{J_2^e}}, \\
\tilde{K}(\phi_2) &=
\frac{J_1^\varepsilon(1 + \nu + \phi_2 E) - 6\alpha\phi_2 E\sqrt{J_2^e}}{3J_1^\varepsilon(1 - 2\nu)(1 + \nu + \phi_2 E)}E.
\label{const_D-P2}
\end{align}
Finally, the multiaxial hardening law for a linear hardening material (cf.\ Fig.~\ref{fig01}) can be obtained from the uniaxial one, Eq.~(\ref{hard}), by substituting $\varepsilon_x^p$ with the equivalent plastic strain, which can be defined for the Drucker-Prager material as
\begin{equation}
\hat{\varepsilon}^p = \frac{\sqrt{3}E}{(1 + \nu^p) E - \nu E^p} \sqrt{J_2^{e^p}} =
\frac{2(\sqrt{3} + 3\alpha)}{3} \sqrt{J_2^{e^p}}.
\end{equation}
This gives
\begin{equation}
\label{ks2}
k_s \left(\sqrt{J_2^{e^p}}\right) = \alpha J_1^\sigma + \sqrt{J_2^s} =
\left( \alpha + \frac{1}{\sqrt{3}} \right) \left( \omega\sqrt{J_2^{e^p}} + \sigma_s \right),
\end{equation}
where
\begin{equation}
\omega = \frac{\sqrt{3}E^pE}{(1 + \nu^p)E - \nu E^p} =
\frac{2E^p(\sqrt{3} + 3\alpha)}{3}.
\end{equation}
In case of the uniaxial stress test discussed in the previous section, Eqs. (\ref{ratio_fi}) and (\ref{D-P_1}) allow us to determine the unknown parameter $\alpha > 0$ from the experiment
\begin{equation}
\label{alpha}
\alpha = \cfrac{J_1^\sigma }{6\sqrt{J_2^s}} \frac{(1 - 2\nu^p)E+2\nu E^p }{(1 + \nu^p)E-\nu E^p} =
\cfrac{1}{2\sqrt{3}} \,\frac{(1 - 2\nu^p)E+2\nu E^p }{(1 + \nu^p)E-\nu E^p}<\cfrac{1}{2\sqrt{3}}.
\end{equation}
Thus, to complete the analysis, the only unknown function $\phi_2$ should be defined. In general, this function depends on both strain invariants $J_1^\varepsilon$ and $J_2^e$
\begin{equation}
\phi_2 = \phi_2(J_1^\varepsilon,J_2^e).
\end{equation}
To this end, note that we can eliminate $J_1^\sigma$ and $J_1^{\varepsilon^p}$ from Eqs.~(\ref{general})$_1$, (\ref{ratio}) and then solve Eqs.~(\ref{general})$_1$, (\ref{ks2}) to obtain $J_2^s(J_1^\varepsilon,J_2^e)$ and $J_2^{e^p}(J_1^\varepsilon,J_2^e)$. The function $\phi_2(J_1^\varepsilon,J_2^e)$ is then derived as
\begin{equation}
\label{phi2dp}
\phi_2(J_1^\varepsilon,J_2^e) = \frac{\sqrt{J_2^s}}{\sqrt{J_2^{e^p}}} =
\frac{3 \alpha (1 + \nu) E J_1^\varepsilon + (1- 2 \nu) \left[3 E \sqrt{J_2^e} - \left(3 \alpha+\sqrt{3}\right) (1 + \nu) \sigma_s\right]}
{E \left\{ 18 \alpha^2 E \sqrt{J_2^e} - 3\alpha[E J_1^\varepsilon - (1 - 2 \nu) (\omega \sqrt{J_2^e}+\sigma_s)] +
\sqrt{3} (1- 2 \nu) (\omega \sqrt{J_2^e}+\sigma_s) \right\}}.
\end{equation}
Finally, the function $\phi_1(J_1^\varepsilon,J_2^e)$ is obtained from (\ref{fi_1})
\begin{equation}
\label{phi1dp}
\phi_1(J_1^\varepsilon,J_2^e) =
\frac{6 \alpha
\left\{3 \alpha (1 + \nu) E J_1^\varepsilon + (1 - 2 \nu) \left[3 E \sqrt{J_2^e} - \left(3 \alpha + \sqrt{3}\right) (1 + \nu) \sigma_s \right]\right\}}
{E
\left[3 E (J_1^\varepsilon - 6\alpha \sqrt{J_2^e}) + \left(3 \alpha + \sqrt{3}\right) (1 + \nu) (\omega J_1^\varepsilon + 6 \alpha \sigma_s)\right]}.
\end{equation}
The evaluation of the functions $\phi_1$ and $\phi_2$ of the two strain invariants $J_1^\varepsilon$ and $J_2^e$ for a Drucker-Prager material having the uniaxial stress-strain curve given in Fig.~\ref{fig01} ($E$ = 813 MPa; $\nu$ = 0.3; $\sigma_s$ = 50 MPa; $E^p$ = 81.3 MPa; $\nu^{ep} = 0.2$; $\nu^p = 0.22$) is shown in Fig.~\ref{fig04}. It is noted that the function $\phi_1$ shows a discontinuity line emanating from a pure shear state on the yield surface and separating regions of different sign. The function $\phi_2$ also shows a discontinuity line emanating from the vertex of the Drucker-Prager cone.
%

The deformation theory developed in this section for Drucker-Prager material with linear hardening has been validated by comparison with the flow theory for several monotonic loadings. Finite element simulations on a single element with displacement control have been performed in ABAQUS and results (not reported here) are always in agreement with the deformation theory.

We present here two examples of monotonic loadings, designed in such a way to investigate the discontinuities shown by the functions $\phi_1$ and $\phi_2$. It will be seen that these discontinuities are genuine features of the Drucker-Prager model with associated flow rule and are obtained also by the more rigorous flow theory of plasticity.

\begin{figure}[!htcb]
\centering
\includegraphics[width=130mm]{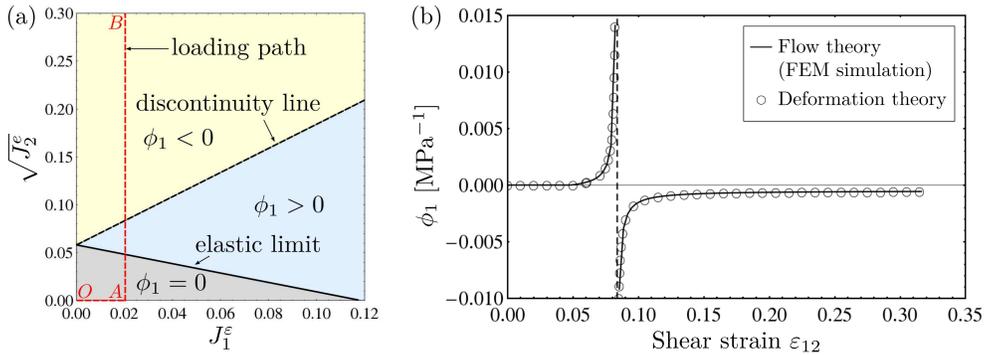}
\caption{(a) Function $\phi_1$ in the $J_1^\varepsilon$-$\sqrt{J_2^e}$ plane: elastic domain $\phi_1 = 0$ (light grey), plastic domain with a discontinuity line separating the regions $\phi_1 > 0$ (light blue) and $\phi_1 < 0$ (light yellow). The red dashed line indicates the loading path. (b) Function $\phi_1$ along the isochoric path $AB$. The discontinuity is clearly visible and results of finite element simulation and deformation theory are in agreement.}
\label{fig_prova_06_a}
\end{figure}

In the first example, the following monotonic loading has been applied in two steps (see Fig.~\ref{fig_prova_06_a}): first a dilatational deformation (from point $O$ to point $A$ in Fig.~\ref{fig_prova_06_a}a), followed by an isochoric deformation (shearing in the plane $x_1$-$x_2$, from point $A$ to point $B$ in Fig.~\ref{fig_prova_06_a}a). This specific loading path has been chosen as it crosses the discontinuity line for the function $\phi_1$. The value of the function $\phi_1$ along the isochoric path $AB$ is shown in Fig.~\ref{fig_prova_06_a}b: results of finite element simulation and deformation theory are in agreement.
%

\begin{figure}[!htcb]
\centering
\includegraphics[width=140mm]{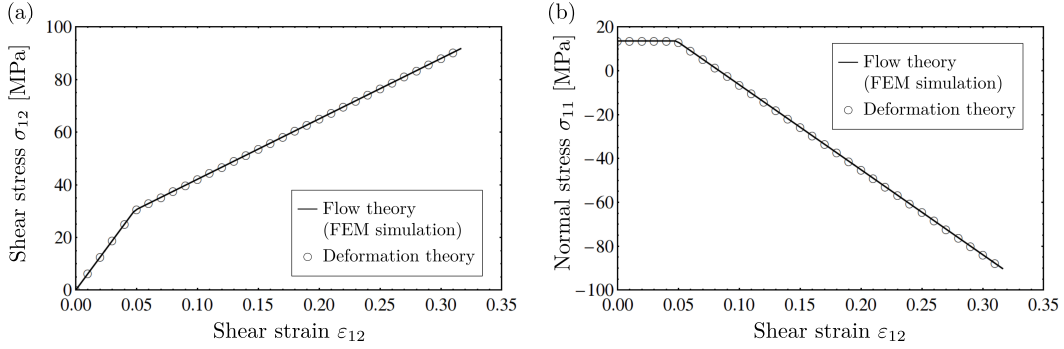}
\caption{Response of the material in terms of stress-strain curves along the isochoric path $AB$ of Fig.~\ref{fig_prova_06_a}a: (a) shear stress $\sigma_{12}$; (b) normal stress $\sigma_{11}$.}
\label{fig_prova_06_b}
\end{figure}

\noindent
In Fig.~\ref{fig_prova_06_b}, we present the stress-strain curves along the isochoric path $AB$ of Fig.~\ref{fig_prova_06_a}a. In particular, Fig.~\ref{fig_prova_06_b}a shows the shear stress $\sigma_{12}$ and Fig.~\ref{fig_prova_06_b}b the normal stress $\sigma_{11}$, both as a function of the shear strain $\varepsilon_{12}$. The agreement between flow theory (finite element simulation) and deformation theory is excellent. It is noted also that the singularity of the function $\phi_1$ does not invalidate the results of the deformation theory model, as the deformation can be continued beyond the discontinuity line.
%

\begin{figure}[!htcb]
\centering
\includegraphics[width=125mm]{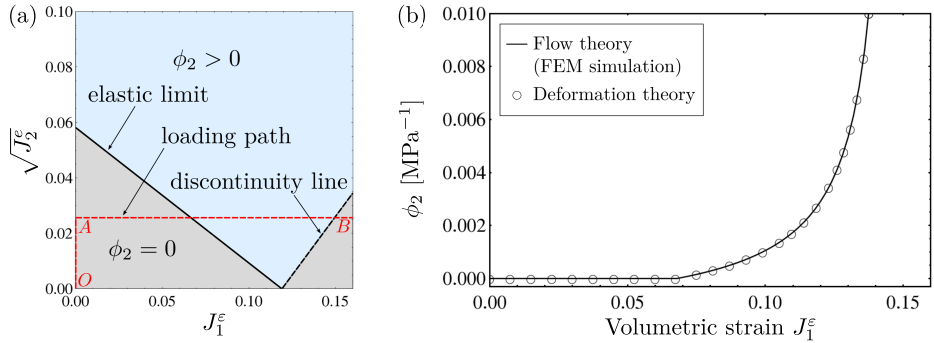}
\caption{(a) Function $\phi_2$ in the $J_1^\varepsilon$-$\sqrt{J_2^e}$ plane: elastic domain $\phi_2 = 0$ (light grey), plastic domain with a discontinuity line separating the regions $\phi_2 > 0$ (light blue) and $\phi_2 = 0$ (light grey). The red dashed line indicates the loading path. (b) Function $\phi_2$ along the dilatational path $AB$. The discontinuity is clearly visible and results of finite element simulation and deformation theory are in agreement.}
\label{fig_prova_08_a}
\end{figure}

In the second example, the following monotonic loading has been applied in two steps (see Fig.~\ref{fig_prova_08_a}): first a isochoric deformation (shearing in the plane $x_1$-$x_2$, from point $O$ to point $A$ in Fig.~\ref{fig_prova_08_a}a), followed by a dilatational deformation (from point $A$ to point $B$ in Fig.~\ref{fig_prova_08_a}a). This specific loading path has been chosen as it crosses the discontinuity line for the function $\phi_2$. The value of the function $\phi_2$ along the dilatational path $AB$ is shown in Fig.~\ref{fig_prova_08_a}b: results of finite element simulation and deformation theory are in agreement.
%

\begin{figure}[!htcb]
\centering
\includegraphics[width=140mm]{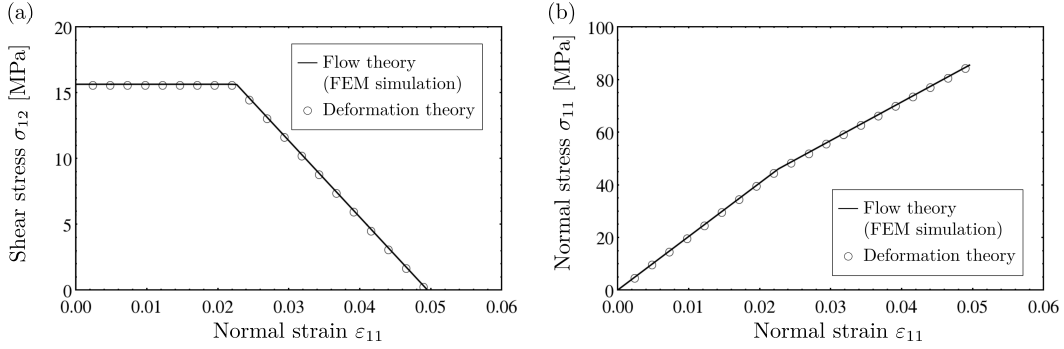}
\caption{Response of the material in terms of stress-strain curves along the dilatational path $AB$ of Fig.~\ref{fig_prova_08_a}a: (a) shear stress $\sigma_{12}$; (b) normal stress $\sigma_{11}$.}
\label{fig_prova_08_b}
\end{figure}

\noindent
In Fig.~\ref{fig_prova_08_b}, we present the stress-strain curves along the dilatational path $AB$ of Fig.~\ref{fig_prova_08_a}a. In particular, Fig.~\ref{fig_prova_08_b}a shows the shear stress $\sigma_{12}$ and Fig.~\ref{fig_prova_08_b}b the normal stress $\sigma_{11}$, both as a function of the dilatational strain $\varepsilon_{11}$. Again, the agreement between flow theory (finite element simulation) and deformation theory is excellent. However, now the deformation cannot be continued beyond the discontinuity line, as the function $\phi_2$ would become negative, which is unphysical. This is due to the fact that the Drucker-Prager model with associate plastic flow always predicts plastic dilatancy.
%


\subsection{Plane strain formulation of transmission conditions}

The two-dimensional plane strain formulation is relevant for the analysis if the conditions of the plane strain are satisfied in the system, i.e. the following components of the strain and stress tensors vanish within the interphase:
\begin{equation}
\label{2D}
\varepsilon_{j3} = 0, \quad \sigma_{i3} = 0, \quad j = 1,2,3, \quad i = 1,2,
\end{equation}
while the remaining components are functions of the independent variables $x_1$ and $x_2$ only.

Finally, the remaining component of the stress tensor can be evaluated from
\begin{equation}
\label{sigma_z}
\sigma_{33} =  (\varepsilon_{11} + \varepsilon_{22}) \tilde{\lambda}(\phi_1,\phi_2).
\end{equation}
Taking into account (\ref{2D}) and the Hooke's law (\ref{hooke}), we obtain the leading terms of the strain tensor invariants \cite{chapter}:
\begin{equation}
\label{strain-as1}
J_1^\varepsilon = \frac{\partial u_2}{\partial x_2} = \frac{\sigma_{22}}{\tilde{\lambda} + 2\tilde{\mu}},
\end{equation}
\begin{equation}
\label{strain-as2}
J_2^e = \frac{1}{3} \left(\frac{\partial u_2}{\partial x_2}\right)^2 + \frac{1}{4} \left(\frac{\partial u_1}{\partial x_2}\right)^2 =
\frac{1}{3} \frac{\sigma^2_{22}}{(\tilde{\lambda} + 2\tilde{\mu})^2} + \frac{1}{4} \frac{\sigma_{12}^2}{\tilde{\mu}^2}.
\end{equation}
Moreover, all the strain and stress components here are functions only of the spatial variable $x_1$. Finally, relationship (\ref{fin1}) remains valid,
while (\ref{fin2}) is rewritten in the form:
\begin{equation}
\label{strain-as4a}
J_2^e = \frac{1}{12h^2}\jump{u_2}^2 + \frac{1}{16h^2}\jump{u_1}^2.
\end{equation}
Note that the transmission conditions (\ref{tc1}) and (\ref{tc2}) are defined now for the two-components vectors $\bt_2 = [\sigma_{21},\sigma_{22}]^T$,
$\bu = [u_1,u_2]^T$ and the matrix $\bA$ is given by
\begin{equation}
\bA =
\begin{pmatrix}
\tilde{\mu} & 0 \\
0 & \tilde{\lambda} + 2\tilde{\mu}
\end{pmatrix}.
\end{equation}
It should be noted that the transmission conditions for plane strain conditions can be written in the compact form as
\begin{equation}
\label{tc2d}
\jump{\sigma_{12}} = 0, \quad\jump{\sigma_{22}} = 0, \quad
\sigma_{12} = F_1(\jump{u_1},\jump{u_2}), \quad \sigma_{22} = F_2(\jump{u_1},\jump{u_2}),
\end{equation}
where the functions $F_1$ and $F_2$ are given by
\begin{equation}
\label{F1F2}
F_1 = \frac{1}{2h} \tilde{\mu}(\phi_1,\phi_2) \jump{u_1}, \quad
F_2 = \frac{1}{2h} (\tilde{\lambda} + 2\tilde{\mu})(\phi_1,\phi_2) \jump{u_2}.
\end{equation}

\section{Validation of the nonlinear transmission conditions for elasto-plastic pressure-dependent interphases}

\begin{figure}[!htcb]
\centering
\includegraphics[width=90mm]{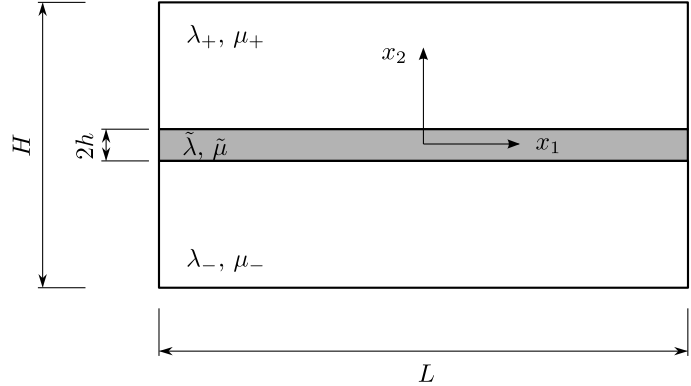}
\caption{Geometry of the three-phase structure.}
\label{fig05}
\end{figure}

The geometry of the considered structure is shown in Fig.~\ref{fig05}. In the analysis the two bonded materials are assumed to be identical and linear elastic, with Young's modulus $E_\pm = 72700$ MPa and Poisson's ratio $\nu_\pm = 0.34$. The geometrical dimensions are $L = 10$ mm, $H = 1$ mm, and $2h = 0.01$ mm. As a result, the small parameter takes the value $\epsilon = 2h/H = 0.01$, which is small enough to justify our asymptotic model.
%

A linear elastic-linear hardening Drucker-Prager model (pressure-dependent yielding) is assumed for the adhesive material within the interphase. The material parameters are defined in Tab.~\ref{tab01} and the corresponding stress-strain curve in uniaxial stress test is shown in Fig.~\ref{fig02}a. For comparison with known results in the literature, a linear elastic-linear hardening von Mises model (pressure-independent yielding) is also considered. All the analyses performed with the von Mises model are in agreement with the literature and are not shown here.
%
\begin{table}[!htcb]
\begin{center}
\begin{tabular}{L{25mm} C{15mm} C{15mm} C{15mm} C{15mm} C{15mm} C{15mm}}
\toprule
Material        & $E$            & $\nu$          & $\sigma_s$       & $E^p$          & $\nu^p$        & $\alpha$       \\
                & (MPa)          &                & (MPa)            & (MPa)          &                &                \\
\midrule
Drucker-Prager  &  813           &  0.3           &  50              &  81.3          &  0.22          &  0.1504        \\
von Mises       &  813           &  0.3           &  50              &  81.3          &  0.53          &  0             \\
\bottomrule
\end{tabular}
\end{center}
\caption{Material parameters used for the validation of transmission conditions.}
\label{tab01}
\end{table}

\noindent
The commercial finite element code ABAQUS is used for the simulation of the mechanical behaviour of the three-phase structure and validation of transmission conditions (\ref{tc2d}). The transmission conditions are based on the deformation theory of plasticity developed in Section \ref{sec03}, whereas the FEM simulations are based on the more general theory of plastic flow. For this reason, only monotonic external loading is considered in the simulations.

\begin{figure}[!htcb]
\centering
\includegraphics[width=110mm]{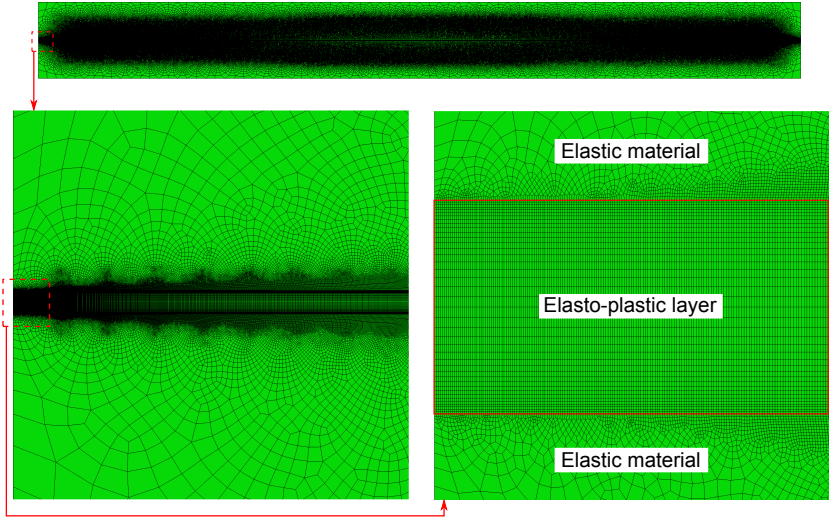}
\caption{Two-dimensional FE-mesh.}
\label{fig06}
\end{figure}

The two-dimensional FE-mesh is shown in Fig.~\ref{fig06}. It has been generated automatically using the mesh generator available within ABAQUS with 4-node bilinear, reduced integration elements with hourglass control (CPE4R). A strong mesh refinement is used within the intermediate elastoplastic layer. The mesh consists of $757774$ nodes and $757599$ elements.

The validation of transmission conditions is performed as follows. From the FEM simulation we obtain the displacements at the top ($x_2 = h$) and bottom ($x_2 = -h$)
of the intermediate layer and evaluate the jumps $\jump{u_1}$ and $\jump{u_2}$. Then, we can compute in sequence: the strain invariants $J_1^\varepsilon$ and
$J_2^e$ from (\ref{fin1}) and (\ref{strain-as4a}), the functions $\phi_1$ and $\phi_2$ involved in the deformation theory from (\ref{phi1dp}) and (\ref{phi2dp}),
the generalized Lam\'e constants $\tilde{\lambda}$ and $\tilde{\mu}$ from (\ref{t_lambda}), and finally the tractions predicted by the transmission
conditions $F_1(\jump{u_1},\jump{u_2})$ and $F_2(\jump{u_1},\jump{u_2})$ from (\ref{F1F2}). The final check consists in comparing the predicted tractions with
those obtained from the FEM simulation.

%


\subsection{Tensile loading with fixed grips}
\label{sec0401}

\begin{figure}[!htcb]
\centering
\includegraphics[width=75mm]{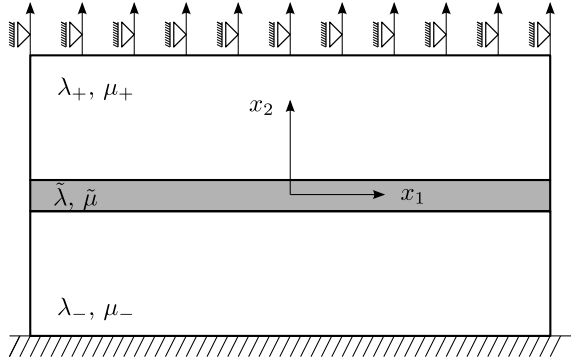}
\caption{Tensile loading with fixed grips.}
\label{fig05b}
\end{figure}

A monotonic tensile loading with fixed grips ($u_1(x_1,H/2) = 0$, $u_2(x_1,H/2) = v_2$) is applied to the top of the bimaterial structure in the range of $v_2/H$ from 0\% to 0.3\% in 100 fixed increments, see Fig.~\ref{fig05b}.
%

\begin{figure}[!htcb]
\centering
\includegraphics[width=150mm]{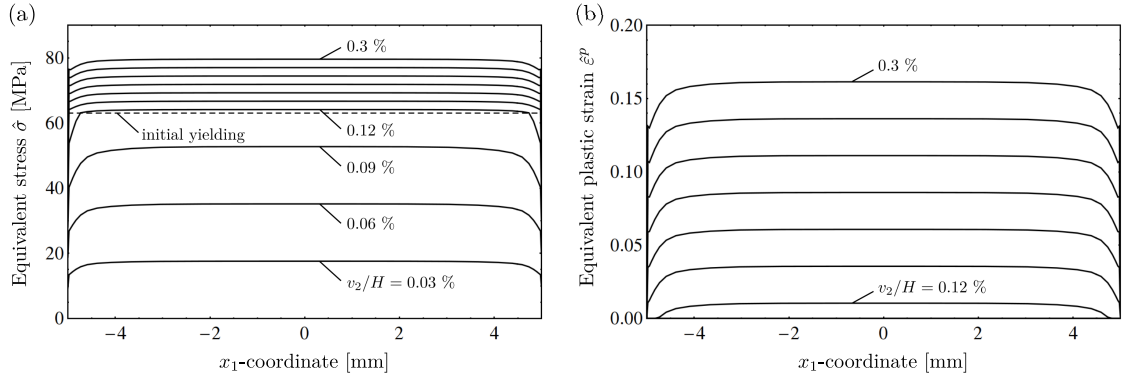}
\caption{Distribution of equivalent stress (a) and equivalent plastic strain (b) along the interphase for different levels of deformation (tensile loading with fixed grips).}
\label{fig10}
\end{figure}

The equivalent stress and the equivalent plastic strain along a horizontal line in the middle of the interphase are shown in Fig.~\ref{fig10}, every ten increments. It is noted that at the 10$^\text{th}$ ($v_2/H = 0.03$\%), 20$^\text{th}$ ($v_2/H = 0.06$\%) and 30$^\text{th}$ ($v_2/H = 0.09$\%) increments the interphase is still in the elastic regime. A visible plastic zone appears in the middle of the interphase at the 40$^\text{th}$ increment ($v_2/H = 0.12$\%). At the 50$^\text{th}$ increment ($v_2/H = 0.15$\%), the interphase is fully plasticised.
%

\noindent
Due to the symmetry of the considered problem (geometry and applied loading), three of the transmission conditions, i.e. $\jump{\sigma_{12}} = \jump{\sigma_{22}} = 0$ and $F_1(\jump{u_1},\jump{u_2}) = \sigma_{12}$, are identically satisfied, because $\jump{u_1} = 0$ and $\sigma_{12} = 0$  hold in this case. The remaining condition $F_2(0,\jump{u_2}) = \sigma_{22}$ has to be verified.

\begin{figure}[!htcb]
\centering
\includegraphics[width=150mm]{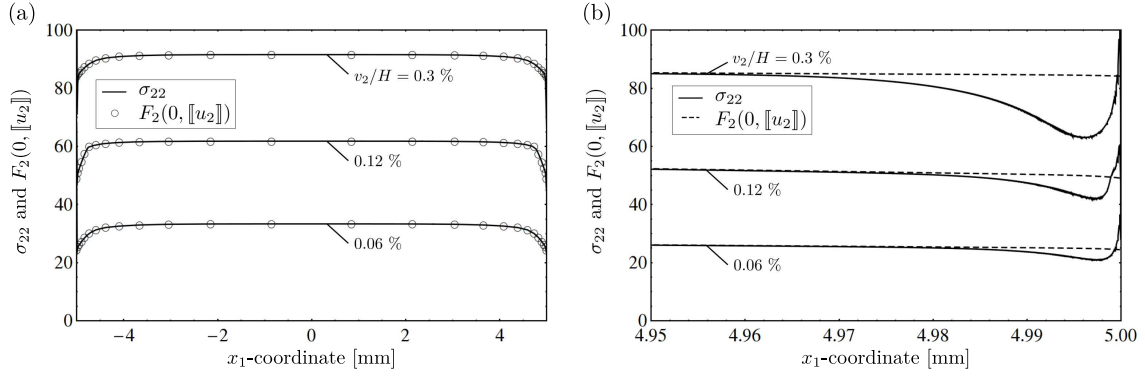}
\caption{(a) Validation of the transmission condition $F_2(0,\jump{u_2}) = \sigma_{22}$ for a pressure-sensitivity elastoplastic interphase (tensile loading with fixed grips). (b) Magnification of the same functions in the interval $4.95 < x_1 < 5$ at the right boundary.}
\label{fig11}
\end{figure}

In Fig.~\ref{fig11}(a), comparisons of the left and right hand sides of the condition $F_2(0,\jump{u_2}) = \sigma_{22}$ are presented. The accuracy of the transmission condition is excellent, both in the elastic and plastic regimes, along the whole interface except for a very small edge zone, as shown in Fig.~\ref{fig11}(b), where a magnification of the same functions in the interval $4.95 < x_1 < 5$ at the right boundary is presented. The size of this edge region is typically 3--4 times the thickness of the interphase.
%

\subsection{Shear loading}
\label{sec0403}

\begin{figure}[!htcb]
\centering
\includegraphics[width=75mm]{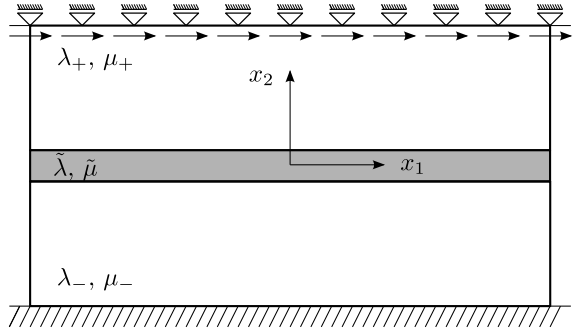}
\caption{Shear loading.}
\label{fig05d}
\end{figure}

A monotonic shear loading ($u_1(x_1,H/2) = v_1$, $u_2(x_1,H/2) = 0$) is applied to the top of the bimaterial structure in the range of $v_1/H$ from 0\% to 0.7\% in 100 fixed increments, see Fig.~\ref{fig05d}.
%

\begin{figure}[!htcb]
\centering
\includegraphics[width=150mm]{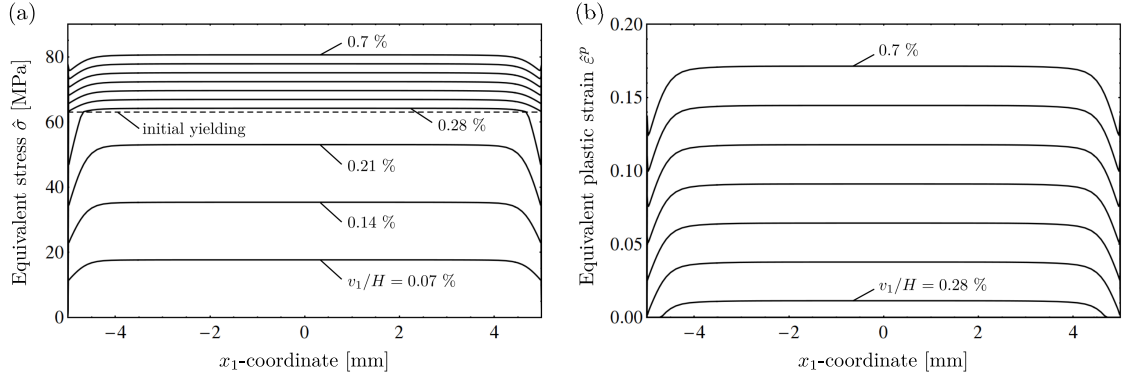}
\caption{Distribution of equivalent stress (a) and equivalent plastic strain (b) along the interphase for different levels of deformation (shear loading).}
\label{fig18}
\end{figure}

The equivalent stress and the equivalent plastic strain along a horizontal line in the middle of the interphase are shown in Fig.~\ref{fig18}, every ten increments. It is noted that at the 10$^\text{th}$ ($v_1/H = 0.07$\%), 20$^\text{th}$ ($v_1/H = 0.14$\%), and 30$^\text{th}$ ($v_1/H = 0.21$\%) increments the interphase is still in the elastic regime. A visible plastic zone appears in the middle of the interphase at the 40$^\text{th}$ increment ($v_1/H = 0.28$\%). At the 50$^\text{th}$ increment ($v_1/H = 0.35$\%), the interphase is fully plasticised.
%

\noindent
Due to the anti-symmetry of the considered problem (geometry and applied loading), two of the transmission conditions, i.e. $\jump{\sigma_{12}} = \jump{\sigma_{22}} = 0$, are identically satisfied. The remaining conditions $F_1(\jump{u_1},\jump{u_2}) = \sigma_{12}$ and $F_2(\jump{u_1},\jump{u_2}) = \sigma_{22}$ have to be verified. In fact, due to the dilatancy of plastic deformation typical of the Drucker-Prager model, we have $\jump{u_2} \neq 0$ and, consequently, tensile stress develops within the interphase, $\sigma_{22} \neq 0$.

\begin{figure}[!htcb]
\centering
\includegraphics[width=150mm]{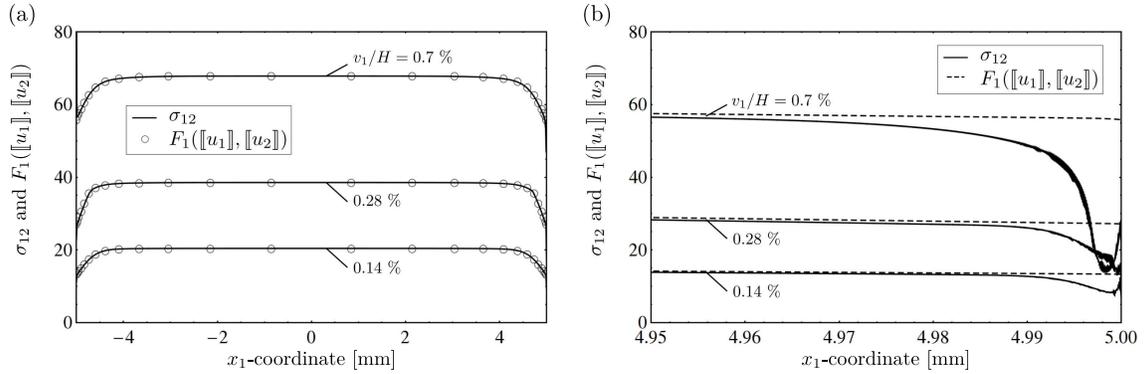}
\caption{(a) Validation of the transmission condition $F_1(\jump{u_1},\jump{u_2}) = \sigma_{12}$ for a pressure-sensitivity elastoplastic interphase (shear loading). (b) Magnification of the same functions in the interval $4.95 < x_1 < 5$ at the right boundary.}
\label{fig19}
\end{figure}

In Fig.~\ref{fig19}(a), comparisons of the left and right hand sides of the condition $F_1(\jump{u_1},\jump{u_2}) = \sigma_{12}$ are presented. The accuracy of the transmission condition is excellent, both in the elastic and plastic regimes, along the whole interface except for a very small edge zone, as shown in Fig.~\ref{fig19}(b), where a magnification of the same functions in the interval $4.95 < x_1 < 5$ at the right boundary is presented. The size of this edge region is typically 3--4 times the thickness of the interphase.
%

\begin{figure}[!htcb]
\centering
\includegraphics[width=150mm]{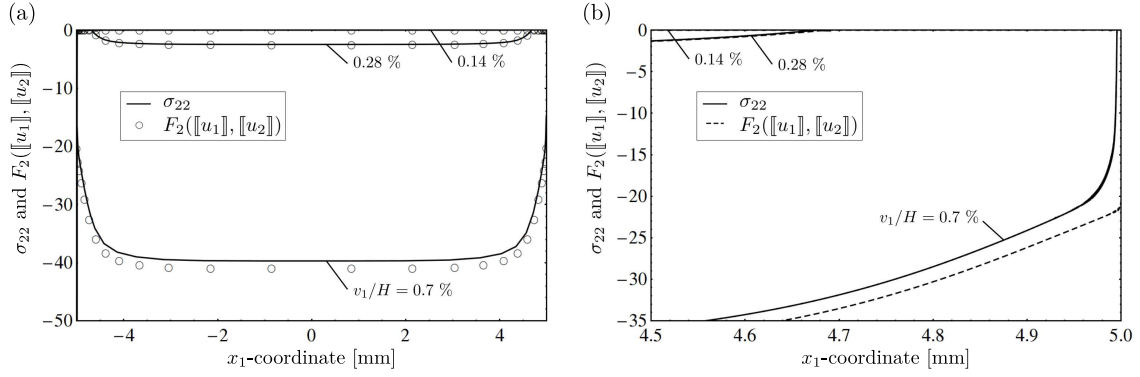}
\caption{(a) Validation of the transmission condition $F_2(\jump{u_1},\jump{u_2}) = \sigma_{22}$ for a pressure-sensitivity elastoplastic interphase (shear loading). (b) Magnification of the same functions in the interval $4.5 < x_1 < 5$ at the right boundary.}
\label{fig20}
\end{figure}

In Fig.~\ref{fig20}(a), comparisons of the left and right hand sides of the condition $F_2(\jump{u_1},\jump{u_2}) = \sigma_{22}$ are presented. In this case, the accuracy of the transmission condition is still excellent in the elastic regime ($v_1/H = 0.14$\%) and in the early stage of plastic deformation ($v_1/H = 0.28$\%). However, at later stages of plastic deformation ($v_1/H = 0.7$\%) the prediction of the transmission condition appears to be less accurate.
%

\begin{figure}[!htcb]
\centering
\includegraphics[width=\textwidth]{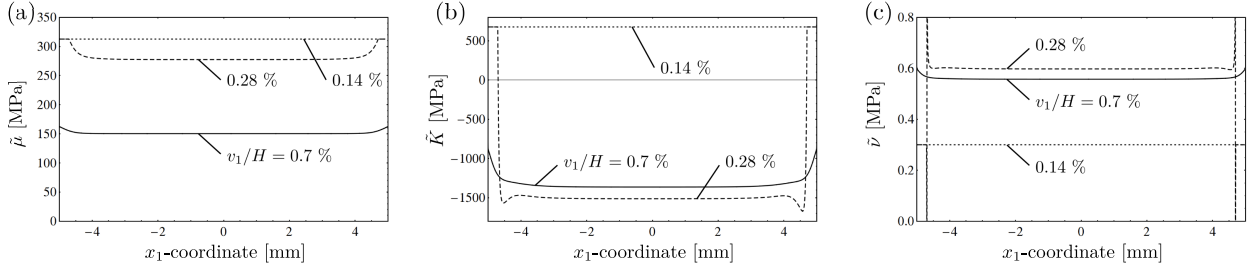}
\caption{Generalized elastic constants $\tilde{\mu}$ (a), $\tilde{K}$ (b), $\tilde{\nu}$ (c), for different levels of deformation (shear loading).}
\label{fig26}
\end{figure}

We observe that, the ellipticity conditions for the generalized elastic constants are not satisfied in the considered stages of plastic deformation ($v_1/H = 0.28$\% and $v_1/H = 0.7$\%), see Fig.~\ref{fig26}. In particular, the generalized bulk modulus $\tilde{K}$ becomes negative and, consequently, the generalized Poisson's coefficient exceeds 0.5. This behaviour is intrinsic of the Drucker-Prager model with associated plastic flow, which always predicts plastic dilatancy.
%

\subsection{Combined loading}
\label{sec0404}

\begin{figure}[!htcb]
\centering
\includegraphics[width=75mm]{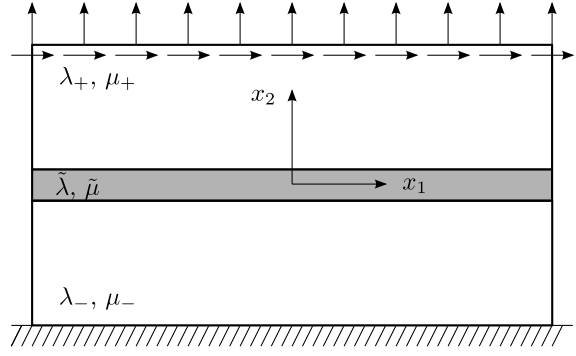}
\caption{Combined loading.}
\label{fig05e}
\end{figure}

A combined loading is applied to the top of the specimen in such a way that the same displacements are prescribed in the $x_1$ as well as the $x_2$ directions ($u_1(x_1,H/2) = u_2(x_1,H/2) = v$) in the range of $v/H$ from 0\% to 0.35\% in 100 fixed increments, see Fig.~\ref{fig05e}.
%

\begin{figure}[!htcb]
\centering
\includegraphics[width=150mm]{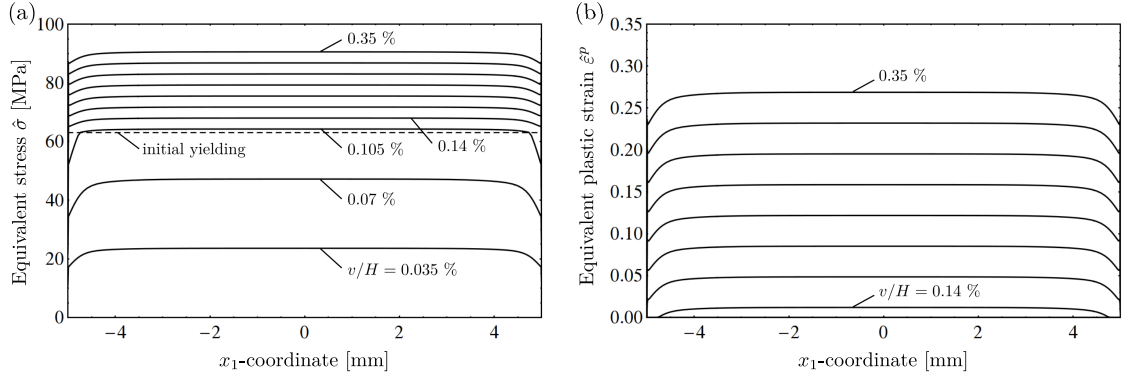}
\caption{Distribution of equivalent stress (a) and equivalent plastic strain (b) along the interphase for different levels of deformation (combined loading).}
\label{fig30}
\end{figure}

The equivalent stress and the equivalent plastic strain along a horizontal line in the middle of the interphase are shown in Fig.~\ref{fig30}, every ten increments. It is noted that at the 10$^\text{th}$ ($v_1/H = 0.06$\%) and 20$^\text{th}$ ($v_1/H = 0.12$\%) increments the interphase is still in the elastic regime. A visible plastic zone appears in the middle of the interphase at the 30$^\text{th}$ increment ($v_1/H = 0.18$\%). At the 40$^\text{th}$ increment ($v_1/H = 0.24$\%), the interphase is fully plasticised.
%

\begin{figure}[!htcb]
\centering
\includegraphics[width=150mm]{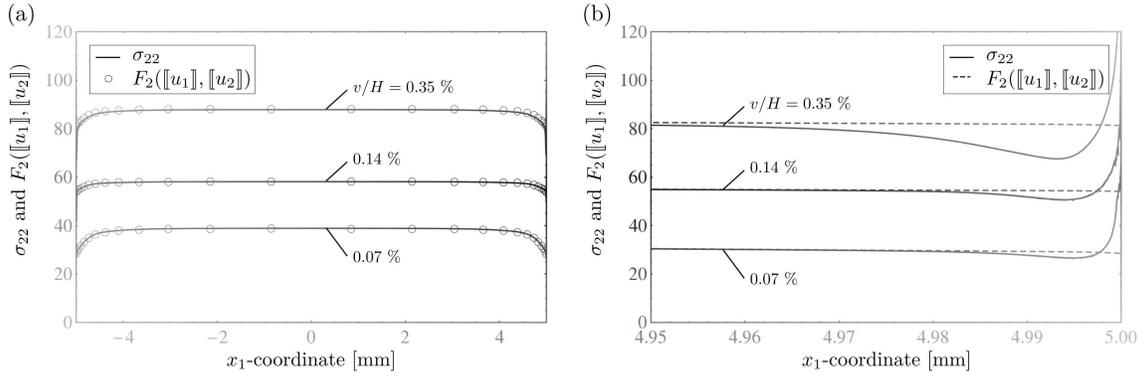}
\caption{(a) Validation of the transmission condition $F_2(\jump{u_1},\jump{u_2}) = \sigma_{22}$ for a pressure-sensitivity elastoplastic interphase (combined loading). (b) Magnification of the same functions in the interval $4.95 < x_1 < 5$ at the right boundary.}
\label{fig32}
\end{figure}

\begin{figure}[!htcb]
\centering
\includegraphics[width=150mm]{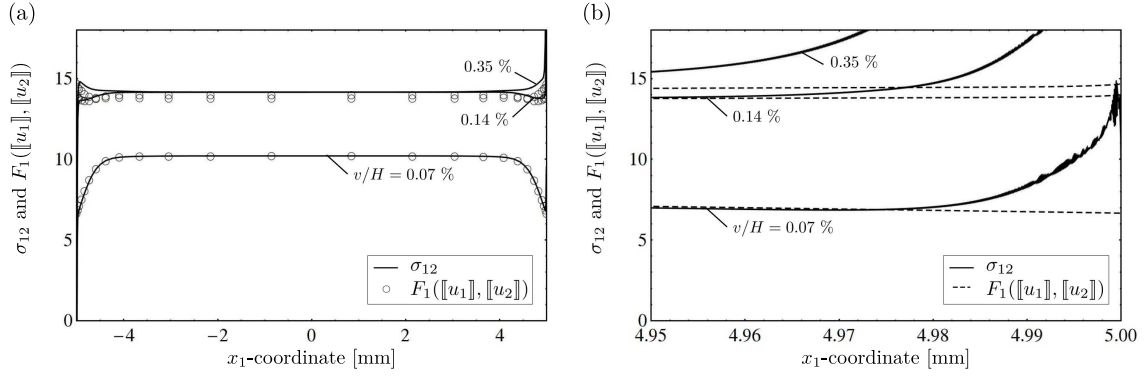}
\caption{(a) Validation of the transmission condition $F_1(\jump{u_1},\jump{u_2}) = \sigma_{12}$ for a pressure-sensitivity elastoplastic interphase (combined loading).(b) Magnification of the same functions in the interval $4.95 < x_1 < 5$ at the right boundary.}
\label{fig31}
\end{figure}

\noindent
For this type of loading, the two conditions $F_1(\jump{u_1},\jump{u_2}) = \sigma_{12}$ and $F_2(\jump{u_1},\jump{u_2}) = \sigma_{22}$ have to be verified.
In Fig.~\ref{fig32}(a), comparisons of the left and right hand sides of the condition $F_2(\jump{u_1},\jump{u_2}) = \sigma_{22}$ are presented. The accuracy of the transmission condition is excellent for all stages of deformation.
In Fig.~\ref{fig31}(a), comparisons of the left and right hand sides of the condition $F_1(\jump{u_1},\jump{u_2}) = \sigma_{12}$ are presented.
%
%

\section{Conclusions}

We have proven that a thin interphase consisting of a soft elasto-plastic pressure dependent material can be effectively modelled by an imperfect nonlinear interface. The latter is governed by nonlinear transmission conditions prescribed along the infinitesimal interface separating two different materials and incorporating the material properties of the adhesive material.
The transmission conditions evaluated here differ essentially from those previously delivered for the case of von Moses plastic law \cite{Ikeda2000,Mishuris2007}. Indeed, the phenomenological functions $\phi_1$ and $\phi_2$ describing the equivalent elasto-plastic interface are not smooth and exhibit a singular behaviour having clear physical sense. Moreover, the conditions evaluated here are universal, in comparison with those delivered previously \cite{chapter}, and can be used for arbitrary stress-state of the considered three-phase body. This makes the evaluated transmission conditions widely applicable and ready to use in FEM analyses (for instance as traction-separation laws in cohesive element formulations) in order to model very thin adhesive layers. This approach avoids extreme mesh refinements usually needed in standard approaches. 

The transmission conditions are validated by numerical examples based on accurate finite element simulations. The accuracy and efficiency of the proposed approach prove to be very high for different monotonic loading conditions. However, as it is the case of any imperfect interface, the conditions are valid along the whole interface apart from a small region close to the sample boundaries where the edge effect is playing a decisive role. To tackle this phenomenon, one needs additionally to use special FEM elements properly modelling the local stress-strain field.
Such strategy (imperfect interface approach enriched by the edge special elements) is an efficient way to compute the details of a structure containing extremely thin soft elasto-plastic interphases of various plastic properties, at least in case of monotonic loading. In case of cyclic load, the range of applicability of the transmission conditions is still to be analysed.

\vspace{10mm}

\noindent
{\bf Acknowledgements}
A.P. gratefully thanks financial support of the Italian Ministry of Education, University and Research in the framework of the
FIRB project 2010 ``Structural mechanics models for renewable energy applications'', G.M. and W.M. gratefully acknowledges the support from the European
Union Seventh Framework Programme under contract numbers PIAP-GA-2011-286110-INTERCER2 and
PIAP-GA-2011-284544-PARM2, respectively.

\bibliographystyle{jabbrv_unsrt}
\bibliography
{%
../../BIBTEX/roaz.bib}

\end{document}

%% file: definitions.tex
\newcommand{\beq}{\begin{equation}}
\newcommand{\eeq}{\end{equation}}

\newcommand{\beqar}{\begin{eqnarray}}
\newcommand{\eeqar}{\end{eqnarray}}
\newcommand{\bit}{\begin{itemize}}
\newcommand{\eit}{\end{itemize}}
\newcommand{\benum}{\begin{enumerate}}
\newcommand{\eenum}{\end{enumerate}}
\newcommand{\barr}{\begin{array}}
\newcommand{\earr}{\end{array}}

\def\XXint#1#2#3{{\setbox0=\hbox{$#1{#2#3}{\int}$}
   \vcenter{\hbox{$#2#3$}}\kern-.5\wd0}}

\def\b0{\mbox{\boldmath $0$}}

\def\be{\mbox{\boldmath $e$}}

\def\bt{\mbox{\boldmath $t$}}
\def\bu{\mbox{\boldmath $u$}}

\def\bA{\mbox{\boldmath $A$}}

\newcommand{\bvarepsilon}{\mbox{\boldmath $\varepsilon$}}

\def\f0{\ensuremath{\mathbb{O}}}

